\documentclass[twocolumn]{aastex631}
\usepackage{apjfonts}
\usepackage{graphicx}
\usepackage{amsmath}
\usepackage{amssymb}
\usepackage{color}

\gdef\xx[#1]{\textcolor{red}{#1}}

\gdef\kms{km\,s$^{-1}$}
\gdef\msun{M$_{\odot}$}

\gdef\lya{Ly\kern 0.09em$\alpha$}
\gdef\ha{H\kern 0.09em$\alpha$}

\newcommand{\GG}[1]{}

\lefthead{van Dokkum et al.}
\righthead{}
\begin{document}

\newcommand\XXX[1]{{\textcolor{red}{\textbf{x\ #1\ x}}}}

\title{Further evidence for a direct-collapse origin of the supermassive black hole at the center of the $\infty$ galaxy}


\author[0000-0002-8282-9888]{Pieter van Dokkum}
\affiliation{Astronomy Department, Yale University, 219 Prospect St,
New Haven, CT 06511, USA}
\affiliation{Dragonfly Focused Research Organization, 150 Washington Avenue, Suite 201, Santa Fe, NM 87501, USA}
\author[0000-0003-2680-005X]{Gabriel Brammer}
\affiliation{Cosmic Dawn Center (DAWN), Niels Bohr Institute, University of Copenhagen, Jagtvej 128, K\o benhavn
N, DK-2200, Denmark}
\author{Connor Jennings}
\affiliation{Astronomy Department, Yale University, 219 Prospect St,
New Haven, CT 06511, USA}
\author[0000-0002-7075-9931]{Imad Pasha}
\affiliation{Astronomy Department, Yale University, 219 Prospect St,
New Haven, CT 06511, USA}
\affiliation{Dragonfly Focused Research Organization, 150 Washington Avenue, Suite 201, Santa Fe, NM 87501, USA}
\author[0009-0005-2295-7246]{Josephine F.\ W.\ Baggen}
\affiliation{Astronomy Department, Yale University, 219 Prospect St,
New Haven, CT 06511, USA}

\begin{abstract}

The $z=1.14$ $\infty$ galaxy consists of two ringed
nuclei with an active supermassive black hole (SMBH) in between them.
The system is likely the result of a nearly face-on collision
between two disk galaxies with massive bulges. In \citet{dokkum:25a}
we suggested that the SMBH may have
formed from shocked and compressed gas at the collision site, in
a runaway gravitational collapse.
Here we test this hypothesis using newly obtained JWST NIRSpec IFU observations.
We first confirm that the system has a cloud of gas
in between the nuclei that is photo-ionized by an AGN-like
object near its center. Next, we constrain the origin of the
SMBH from its radial velocity. If it formed in the cloud its
velocity should be similar to the surrounding gas, whereas it
would be offset if the SMBH had escaped from one of the nuclei or
were associated with a faint galaxy. We find that 
the radial velocity of the SMBH is within $\sim 50$\,\kms\
of that of the surrounding gas, as expected if the SMBH formed within the cloud.
Unexpectedly, we find that both nuclei have active SMBHs as well,
as inferred from very broad H$\alpha$ emission with FWHM\,$\sim 3000$\,\kms.
This rules out scenarios where the central SMBH was ejected from one
of the nuclei in a gravitational recoil.
Taken together, these results strengthen the hypothesis that the object
at the center of the $\infty$ galaxy is a newly formed SMBH.

\end{abstract}


\section{Introduction}

Although
supermassive black holes (SMBHs) reside in the centers of nearly all luminous
galaxies \citep{magorrian:98}, their origins are not well understood
\citep{volonteri:10,regan:24}.
Two of the leading models are that
they began as the $\sim 10^{1-3}$\,\msun\ collapsed remnants of
Population III stars \citep[`light seeds'; see, e.g.,][]{madau:01},
or that they formed in a direct collapse of pre-galactic $\sim 10^{4-5}$\,\msun\ gas clouds
\citep[`heavy seeds'; see][]{haehnelt:93,bromm:03,lodato:06}.

Constraining black hole formation models is notoriously difficult.
The initial stages occur in the centers of assembling galaxies,
and in these complex environments it is difficult to determine how long a
SMBH has been present and what its mass was when it formed.
Besides this conceptual problem
there is a practical barrier: in most models the initial collapse takes
place at $z>15$, and is beyond even the reach of the James Webb Space Telescope (JWST).
In the absense of direct information, constraints so far have been indirect:
by estimating SMBH masses in galaxies at $z=6-10$ one can work
backwards through plausible merger and accretion trees to estimate the initial
mass \citep[see, e.g.,][]{natarajan:24}.
Early JWST studies have found surprisingly high black hole masses in many
galaxies in this redshift range, in qualitative support of heavy seed models
\citep{natarajan:17,natarajan:24,greene:24,matthee:24}.
In this overall context, finding a SMBH just after it formed would be highly informative:
it would directly demonstrate the viability of heavy seed models
and it would constrain simulations of the collapse process.

\begin{figure*}[ht]
  \begin{center}
  \includegraphics[width=0.95\linewidth]{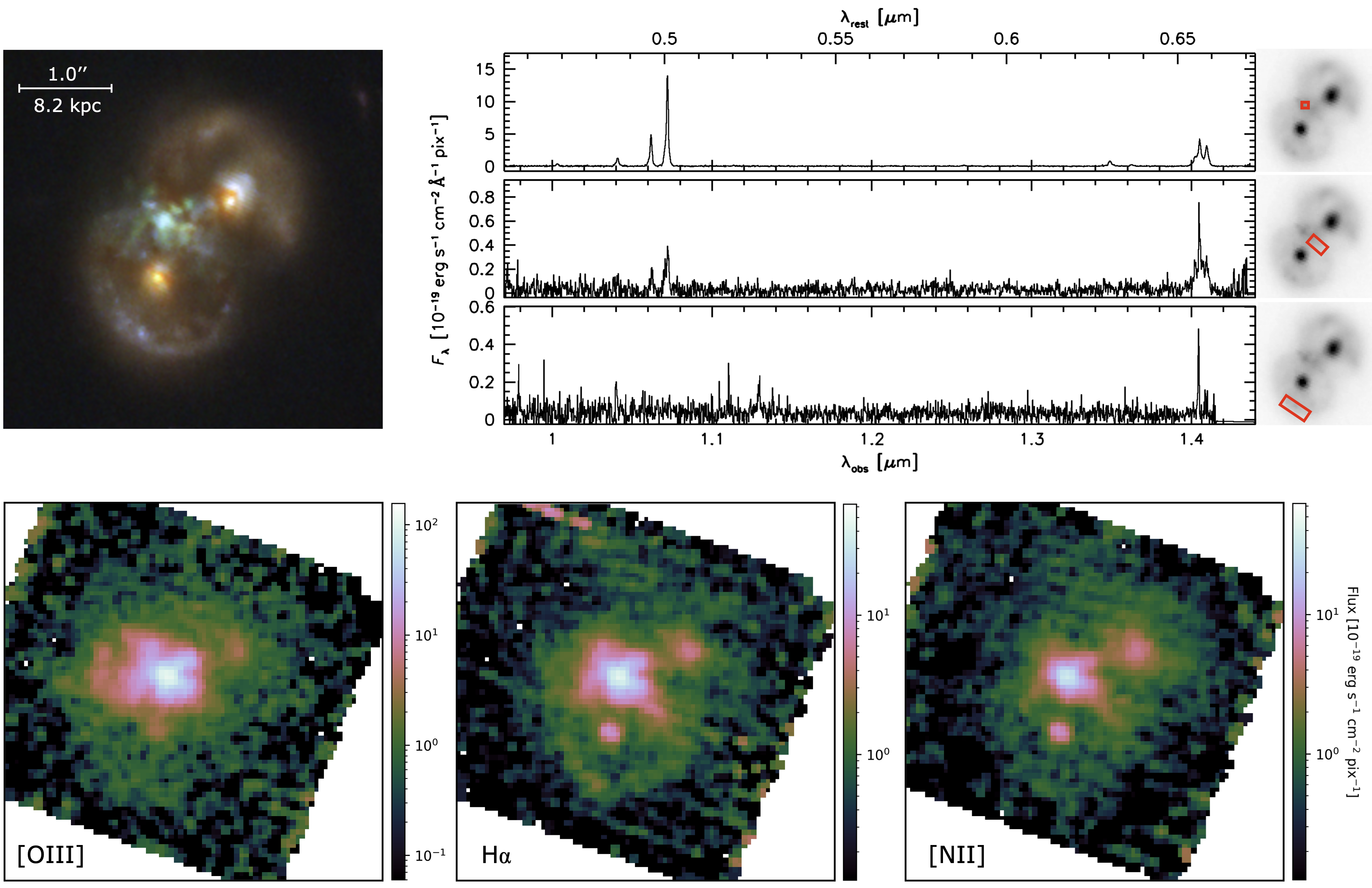}
  \end{center}
\vspace{-0.2cm}
    \caption{
NIRSpec IFU observations of the $\infty$ galaxy.
{\em Top left:} Broad band image of the galaxy created from the NIRCAM F090W (blue),
F115W\,+\,F150W (green), and F200W (red) data, sampled at $0\farcs 02$
resolution with North up and East to the left.
{\em Top right:} Example spectra
from three different regions, highlighting the diversity of spectra in the data cube.
{\em Bottom:} Maps of [O\,III], H$\alpha$, and [N\,II] emission, sampled
with $0\farcs 05$ pixels. There is a cloud
of ionized gas in between the nuclei. Both nuclei, and the SE ring, are detected
in H$\alpha$.
}
\label{ifu_overview.fig}
\end{figure*}

Recently we identified a candidate for such an object \citep[][hereafter paper I]{dokkum:25a}.
It resides in a galaxy at $z=1.14$ with
a total stellar mass of $M_{\rm stars}\sim 3 \times 10^{11}$\,\msun.
The rest-frame near-IR light of the galaxy is dominated by two compact
nuclei with a projected separation of $10$\,kpc. Both nuclei have
a prominent ring or shell around them, giving the system the appearance of a figure eight or
an $\infty$ symbol (see Fig.\ 1). The $\infty$ galaxy is a textbook example of a binary collisional ring
system, where two bulge\,+\,disk galaxies experienced a high speed, nearly head-on
collision. The bulges of the galaxies
mutually disturb the disks, leading to the formation of collisional rings around both of them. The
prototypical example is the galaxy II\,Hz\,4 at $z=0.04$, simulated by \citet{lynds:76}.

The $\infty$ galaxy has an active SMBH with an
X-ray luminosity\footnote{In paper I the X-ray luminosity was erroneously given as $1.5\times 10^{44}$\,erg\,s$^{-1}$, due to an error in the Chandra exposure time \citep[the 16\,ks of the wide COSMOS survey was used, instead of the 162\,ks that is appropriate for the central region; see][]{elvis:09}.} of $L_{\rm X}\sim 1.5\times 10^{43}$\,erg\,s$^{-1}$ and
a radio luminosity of $L_{144\,{\rm MHz}} \sim 2 \times 10^{26}$\,W\,Hz$^{-1}$.
Based on the centroids of a VLA\,3\,GHz map, a Chandra X-ray image,
and the [Ne\,III] line in a Keck spectrum, we determined in paper I that the SMBH is
located in between the two nuclei.
It is one of the best examples of a SMBH that is outside of the center of
a galaxy \citep[see, e.g.,][for other candidates]{civano:10,
dokkum:23,uppal:24}.

As discussed in paper I,
there are three plausible explanations for the presence of a SMBH in such an unusual location.
The first is that it is in a separate galaxy that is too faint to be detected against the glare
of its active nucleus and the rings of the $\infty$ system.
The second is that it separated from its
former host: it could
have been ejected from one of the nuclei after a merger, through gravitational recoil \citep{lousto:11}
or a three-body interaction \citep{bekenstein:73,saslaw:74,hoffman:07}, or it could be
a `wandering' SMBH that has lost its host galaxy through severe tidal stripping \citep{tremmel:18}.

The third explanation is that it was born where it is,
in a runaway gravitational collapse of a dense gas clump during the
collision of the two disk galaxies. The main evidence for this scenario is that the
SMBH appears to be embedded in an extended distribution of ionized
gas, as determined from excess NIRCAM F150W emission. This gas
was likely shocked and compressed during the collision, in a process akin to the bullet cluster
\citep{clowe:06} but on much smaller scales.
Although no simulations of black hole formation in such
environments have yet been performed, star formation is likely suppressed and the formation of
massive clumps may be promoted \citep{silk:19,yeager:19,lee:21,appleton:22}.
Furthermore, high resolution simulations have shown that a runaway collapse
can occur in other extreme environments, such as collapsing pregalactic halos \citep{wise:19}
and gas-rich merger remnants \citep{mayer:10,mayer:15}.

These scenarios predict distinctly different relationships between the SMBH and the surrounding gas.
In the `galaxy origin' scenarios the SMBH is moving through the cloud, at a velocity that is
either characteristic of the mass of the system ($\sim 350$\,\kms) or, in the case of ejection,
exceeds the escape velocity of the nuclei
($\gtrsim 1200$\,\kms). In the direct-collapse scenario the gas is the black hole's birth cloud,
and the SMBH should be approximately at rest with respect to it.
In this {\em Letter}, we constrain the origin of the SMBH in the $\infty$ galaxy
by measuring its velocity  with respect to the surrounding gas.

\section{Data}

\subsection{Observations}

The $\infty$ galaxy was observed with the JWST NIRSpec IFU on May 23, 2025 in
Director's Discretionary program 9327 (PI: van Dokkum). The $\infty$ galaxy nearly fills the
$3\arcsec \times 3\arcsec$ field of view of the IFU. To facilitate sky subtraction, two
connected visits were obtained: one centered on the galaxy, and one in a nearby empty sky
region with identical instrumental settings and exposure time.

The G140H filter 
in combination with the F100LP filter gives a resolving power of $R \approx 2700$ and
wavelength coverage of 0.97\,$\mu$m -- 1.89\,$\mu$m. The detector gap is between $1.437\,\mu$m
and $1.460\,\mu$m for the center of the IFU, which means that the redshifted
[S\,II] doublet was not consistently observed. The NRSIRS2RAPID readout pattern was used, with 40
groups per integration. Four integrations were obtained,
using a standard 4-point dither pattern. This pattern reduces the impact of bad
pixels and provides sampling on a $2\times$ finer pixel grid. The exposure time was
2393\,s for the science exposure and 2393\,s for the offset background exposure.
The total program time, including overhead, was 2.8\,hrs.

\subsection{Data reduction}

The data were reduced using the TEMPLATES data reduction pipeline, as described in \cite{TEMPLATES}.
TEMPLATES is built upon the standard JWST Science Calibration Pipeline \citep{JWSTpipe};
we used version 1.18.0. A brief description of the pipeline follows.
Uncalibrated files generated by the JWST Science Data Processing subsystem (version 2025\_2) were
downloaded from MAST. Detector level processing (JWST stage 1) was performed on the uncalibrated files.
NSClean \citep{rauscher:24} was used to correct $1/f$ noise in both the science and background frames.
The JWST stage 2 spectroscopic pipeline was then run on both science and background observations. No
background subtraction was performed at this stage. The most extreme outlier spaxels (from cosmic rays)
in the stage 2 data were flagged by setting a maximum good value slightly greater than the brightest
emission lines, in both the NRS1 and NRS2 detectors. The NRS1 and NRS2 data from all dithers were then
combined, and background subtraction was performed using the default method in the JWST stage 3
spectroscopic pipeline. This method combines all spaxels in the background data cube
to create a global background, which was then subtracted from each spaxel in the science cube.

Taking
advantage of the subpixel sampling of the dither pattern, the
data were combined onto two pixel scales: $0 \farcs 1$, corresponding to the native
pixel scale, and $0\farcs 05$, corresponding to the angular scale that is sampled by the dither pattern.
The galaxy is not exactly centered in the field of view, as no pointing acquisition was done.
The two cubes were aligned with the images of paper I by
integrating them in the F150W filter and determining their spatial offsets with respect
to the astrometrically-verified NIRCAM F150W image.

\section{A cloud of photo-ionized gas in between the two nuclei}

\subsection{Emission Line Maps and Equivalent Widths}

Emission line maps for the three brightest lines, [O\,III]\,$\lambda5007$, H$\alpha$, and
[N\,II]\,$\lambda6583$, are shown in Fig.\ \ref{ifu_overview.fig}. The maps were created
from the $0\farcs 05$ cube by
summing $\pm 6$ channels, corresponding to $\pm 14$\,\AA\ or $\pm 300$\,\kms\ at H$\alpha$,
centered on
the channels corresponding to the redshifted lines. Continuum images were
created by taking the median over line-free regions. These images were multiplied
by 13 and subtracted from the line maps.
There is some striping in the cube due to
alternating column noise (a known artifact, caused by offsets in the two amplifiers).\footnote{The stripes
can be removed quite effectively by doing
a per-spaxel subtraction of the background cube, but this would lead to an increase in the noise.}

The line maps confirm the existence of a spatially-extended cloud of ionized gas in between the two
nuclei, as was previously inferred from an excess of emission in the F150W NIRCAM filter (see paper I).
The maps show a clear peak, where the flux in the brightest line, [O\,III], reaches
$F_{\rm pix}({\rm [O\,III]}) \approx 1.1\times 10^{-17}$\,erg$^{-1}$\,s$^{-1}$\,cm$^{-2}$\,pix$^{-1}$. 
The total [O\,III] flux in the cloud between the two nuclei is $F({\rm [O\,III]})
=(1.1\pm 0.1) \times 10^{-15}$\,erg$^{-1}$\,s$^{-1}$\,cm$^{-2}$. This is $\sim 5\times$ higher
than was inferred in paper I from a Keck LRIS spectrum. This difference is likely
caused by a combination of slit losses and
calibration errors: the sensitive of Keck/LRIS is very low at $1.07\mu$m
and was calibrated using the fluxes of sky
emission lines (see paper I).
The JWST-measured flux corresponds to a cloud luminosity
of $L({\rm [O\,III]}) = (8\pm 1) \times 10^{42}$\,erg\,s$^{-1}$.
The [O\,III] luminosity alone places the cloud
firmly in the AGN regime:  
normal star forming galaxies typically have $L({\rm [O\,III]}) \lesssim 10^{41}$\,erg\,s$^{-1}$
\citep{kauffmann:03,brinchmann:04}.

There is remarkably little continuum emission in the cloud region. The rest-frame
equivalent width of the [O\,III] line,
as measured directly from the spectra, ranges from $\sim 400$\,\AA\ to $\sim 1600$\,\AA.
The equivalent width is even be higher when projection effects are taken into account:
as shown in paper I the stellar rings around the nuclei are projected onto the central region,
boosting the observed continuum emission. The IFU observations thus confirm the existence of a
cloud of ionized gas in between the two nuclei,
with a luminosity exceeding that of star forming galaxies and very little associated
continuum emission.

\begin{figure*}[ht]
  \begin{center}
  \includegraphics[width=0.85\linewidth]{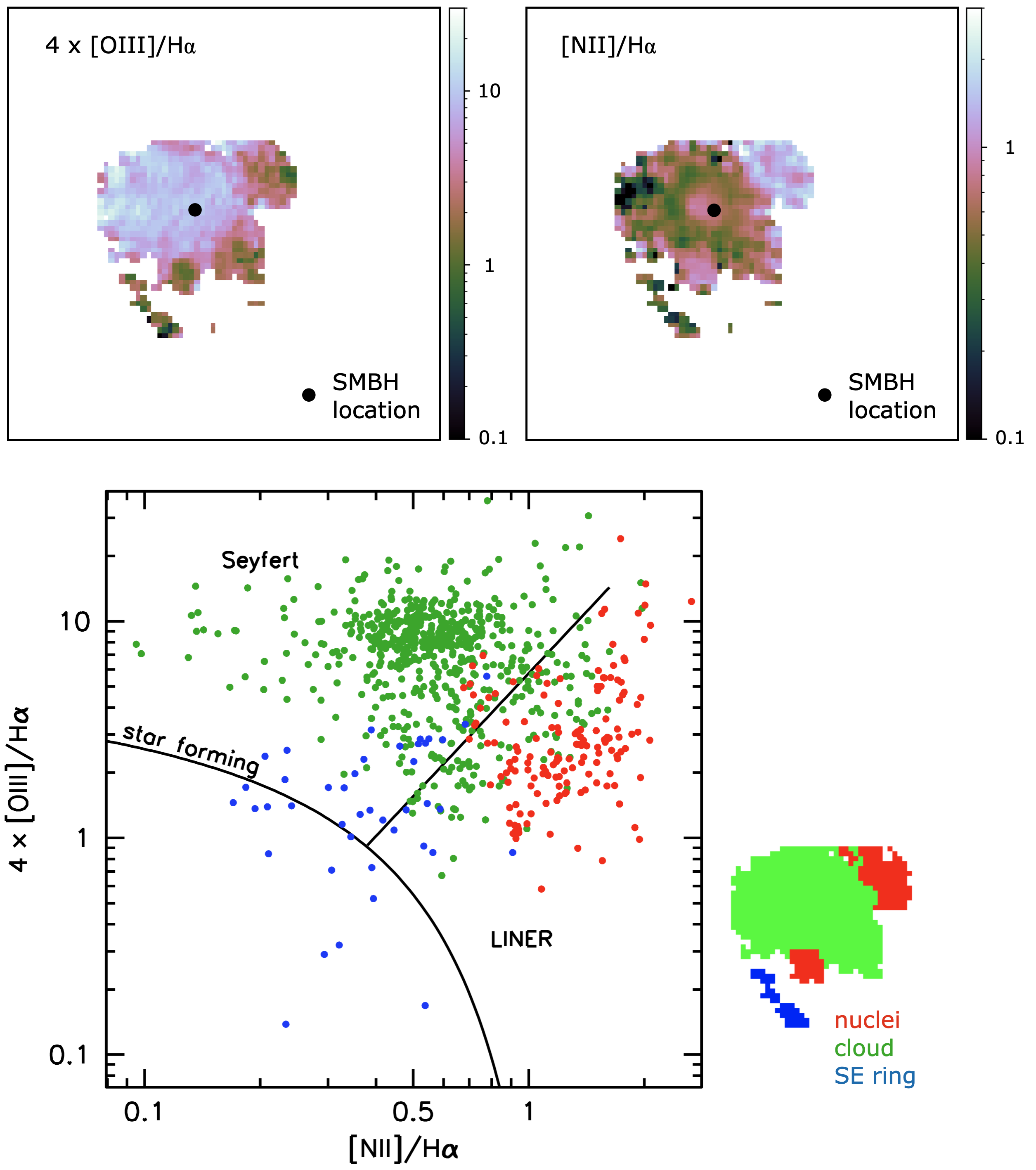}
  \end{center}
\vspace{-0.2cm}
    \caption{
Line ratios. {\it Left:} Map of the [O\,III]/H$\alpha$ ratio, multiplied by a factor of 4
so values can be more easily compared to [O\,III]/H$\beta$ measurements.
{\it Middle:} Map of the [N\,II]/H$\alpha$ ratio. {\it Right:} Individual $0\farcs 05\times 0\farcs 05$
spaxels in the BPT diagram. The cloud shows very high $4\times$\,[O\,III]/H$\alpha$ ratios of $\sim 10$,
and is in the Seyfert regime. The nuclei are in the LINER part of the BPT diagram, and the
ring is consistent with star formation.
}
\label{line_ratios.fig}
\end{figure*}

\subsection{Line Ratios}
\label{lineratios.sec}

The morphology of the $\infty$ galaxy is quite different in the three line
maps shown in Fig.\ \ref{ifu_overview.fig}.
Three representative spectra are shown at top right, extracted from the blue (NRS1) side
of the $0\farcs 05$ data cube. The top spectrum is for the brightest area, near the SMBH;
the middle spectrum is for a region of relatively low
surface brightness in between the nuclei; and the bottom spectrum is extracted from the
SE ring. These spectra highlight the stark differences in the intensities,
ratios, and widths of lines
in various regions of the $\infty$ system.

Line ratio maps, created by dividing the continuum-subtracted
emission line maps, are shown in Fig.\ \ref{line_ratios.fig}. 
The right panel of Fig.\ \ref{line_ratios.fig} shows
individual spaxels in the BPT diagram \citep{baldwin:81}, with the
star forming sequence and the separation between Seyferts and LINERs taken from
\citet{kewley:13} for $z\sim 1$.
We use [O\,III]/H$\alpha$
in lieu of the standard [O\,III]/H$\beta$ ratio, as H$\beta$ is too faint for measuring
reliable line ratios outside of the central regions. For convenience, we multiply
the [O\,III]/H$\alpha$ ratio by 4 so it is on the approximate scale of
[O\,III]/H$\beta$ ratios. In the central $0\farcs 4\times 0\farcs4$
the average observed ${\rm H}\alpha/{\rm H}\beta$ ratio is $\approx 4.4$.\footnote{This H$\alpha/{\rm H}\beta$ ratio corresponding to $A_V \approx
1.2-1.5$\,mag for intrinsic ratios of $2.9-3.1$ \citep{gaskell:84}
and a \citet{calzetti:00} attenuation law.
The attenuation is similar to that derived from SED fitting of the two nuclei, and qualitatively
consistent with the dust lanes that are visible throughout the $\infty$ system (see paper I).}

The [O\,III]/H$\alpha$ and [N\,II]/H$\alpha$
ratios show clear patterns. The cloud between the nuclei is highly ionized throughout,
with $4\times{\rm [O\,III]/H}\alpha \sim 10$. The [N\,{\sc ii}]/H$\alpha$ ratios are in the range
$0.3-0.8$, with the highest values near the SMBH. The gas near the two nuclei falls in the LINER region
of the BPT diagram, with [N\,{\sc ii}]/H$\alpha \gtrsim 1$ and relatively low [O\,III]/H$\alpha$.
The only gas in the $\infty$ system exhibiting ionization typical of H\,II regions is in the
SE ring. This is consistent with the colors and morphology of the galaxy in NIRCAM images:
there are blue star forming complexes in the SE ring (see Fig.\ \ref{ifu_overview.fig}),
and this region is the brightest part of the galaxy after the
central source in the rest-frame far-UV (see paper I).

The spectra show several other emission lines in the brightest region, near the SMBH: we detect
He\,II\,$\lambda 4686$, He\,I\,$\lambda 5876$, the [O\,I]\,$\lambda\lambda 6300,6364$ doublet,
the [O\,II]\,$\lambda\lambda 7319,7330$ doublet, and [Ar\,III]\,$\lambda 7135$.
A more comprehensive analysis of emission lines in the cloud will be performed in a future study.
We note here that the strongest constraint on the ionizing energy comes from the [Ne\,V]\,$\lambda\lambda 3426,3346$
doublet detected with LRIS (see paper I), which requires photons with energies $\geq 97.11$\,eV
($\lambda\leq 128$\,\AA).

\subsection{Ionization Mechanism and Geometry of the Cloud}
\label{photo.sec}

The locations of the cloud spaxels in the BPT diagram rule out star formation as the
dominant ionization mechanism.
Star formation is also inconsistent with the [O\,III] luminosity of the cloud, 
the extreme equivalent width of the lines, and the lack of visible star forming regions in the NIRCAM images.
It is more difficult to differentiate shocks from photoionization by an active SMBH. Shocks alone generally
produce low ionization lines, but the shock front can have a photoionized precursor that can
mimic an AGN spectrum \citep[see][]{dopita:95}.

\begin{figure}[ht]
  \begin{center}
  \includegraphics[width=1.0\linewidth]{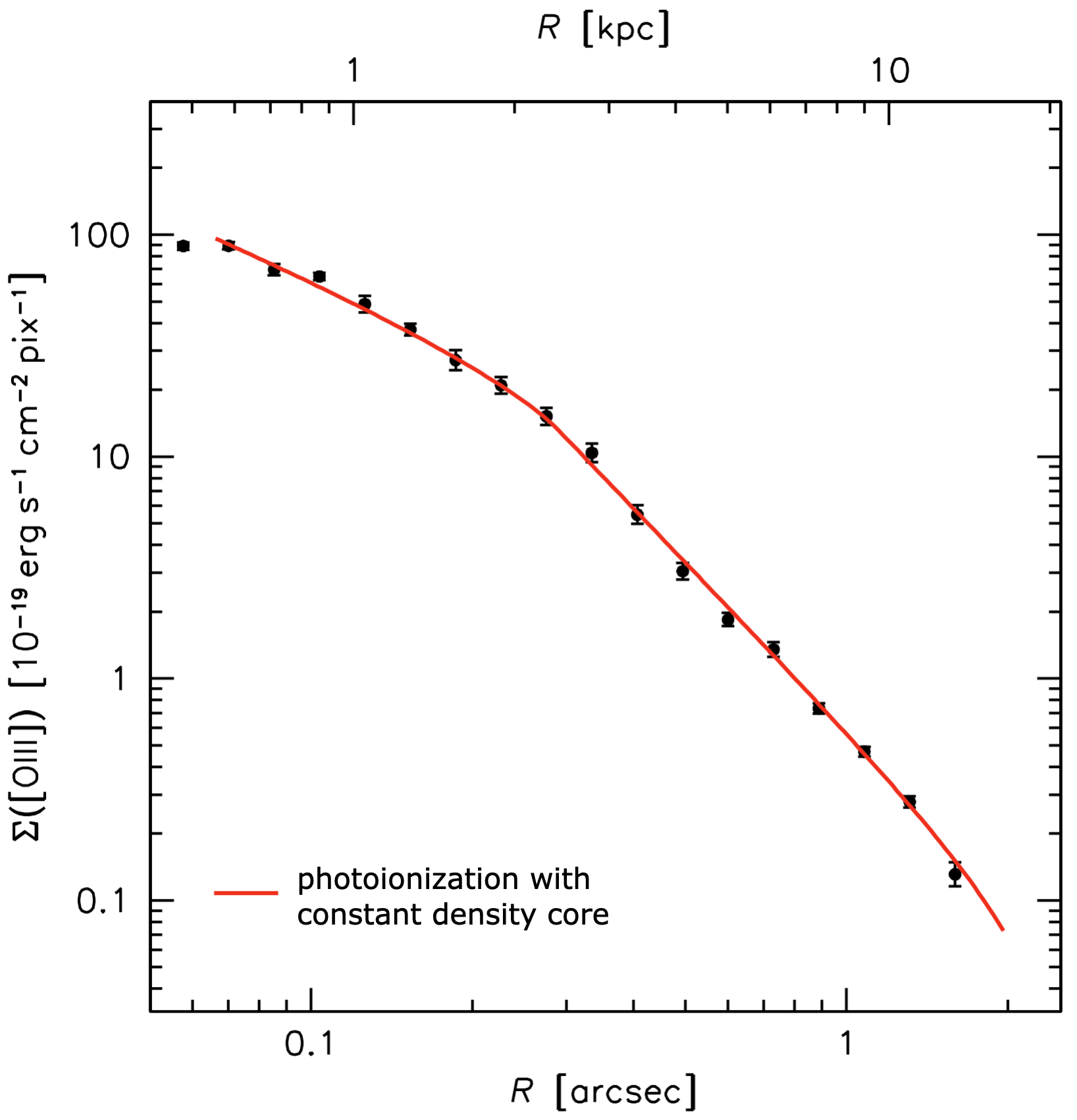}
  \end{center}
\vspace{-0.2cm}
    \caption{
Radial average surface brightness profile of the [O\,III] emission in the cloud.
The profile has a clear peak and falls off monotonically.
The red line shows the expectation for photoionization, with the gas having a constant
density core with a radius of $r_{\rm core}=2.5$\,kpc and a powerlaw distribution with $n
\propto r^{-0.75}$ at $r>r_{\rm core}$.
}
\label{radprof.fig}
\end{figure}

There are two reasons why shock models are disfavored. First, the [O\,III] surface brightness is extremely
high. The average attenuation-corrected surface brightness in the cloud is $\Sigma({\rm [O\,III]})\sim 4\times 
10^{41}$\,erg\,s$^{-1}$\,kpc$^{-2}$, with  $\Sigma({\rm [O\,III]})
\sim 1.5 \times 10^{42}$\,erg\,s$^{-1}$\,kpc$^{-2}$ in the central kpc. This can be
compared to typical surface
brightnesses in shock\,+\,precursor models of $10^{38}$ -- $10^{40}$\,erg\,s$^{-1}$\,kpc$^{-2}$ \citep{allen:08}.
Second, the spatial distribution of the line emission clearly points to photoionization by
a central object. 
The surface brightness profile of the cloud, centered on the peak, is shown in Fig.\ \ref{radprof.fig}.
The surface brightness decreases monotonically with distance from the peak, as expected
in photoionization models. In shock models, by contrast, the surface brightness distribution is
typically complex along the front,
as there is no central source and
the ionization reflects the localized interaction of the shock with the ambient gas
\citep[see, e.g.,][]{tilak:05,privon:08}.

Under the assumption of photoionization by a central object, we can use the surface brightness
profile to constrain the density distribution of the cloud.
We assume that the gas is optically thin in the [O\,III] line,
isotropically photoionized by a central source,
resulting in an ionizing photon flux that declines as $\Phi(r) \propto 1/r^2$,
and isothermal, with constant excitation and ionization conditions throughout.
We also assume spherical symmetry.
The  local [O\,III] emissivity at radius $r$ is then given by:
\begin{equation}
\epsilon(r) \propto n_e^2(r)\Phi(r) \propto \frac{n^2(r)}{r^2}.
\end{equation}

Motivated by the $\sim R^{-1}$ slope of the observed profile out to $\sim 0\farcs 3$
we assume that the gas density follows a core\,+\,power-law profile,
\begin{equation}
n(r) =
\begin{cases}
n_0 & \text{if } r < r_{\mathrm{core}} \\
n_0 \left( \dfrac{r}{r_{\mathrm{core}}} \right)^{-\alpha} & \text{if } r \geq r_{\mathrm{core}},
\end{cases}
\end{equation}
which leads to an emissivity profile
\begin{equation}
\epsilon(r) =
\begin{cases}
\dfrac{n_0^2}{r^2} & \text{if } r < r_{\mathrm{core}} \\
\dfrac{n_0^2}{r_{\mathrm{core}}^{-2\alpha}} \cdot r^{-2\alpha - 2} & \text{if } r \geq r_{\mathrm{core}}.
\end{cases}
\end{equation}

The observed surface brightness $\Sigma(R)$ at projected radius $R$ is the line-of-sight integral of the emissivity:
\begin{equation}
\Sigma(R) = 2 \int_0^{\sqrt{r_{\mathrm{max}}^2 - R^2}} \epsilon\left( \sqrt{R^2 + l^2} \right) \, dl.
\end{equation}
This expression is evaluated numerically.
Besides the overall normalization the model
has three free parameters:
$r_{\mathrm{core}}$, the 3D radius of the constant density core, 
$\alpha$, the slope of the outer density profile, and
$r_{\rm max}$, the outer edge of the cloud.
The model is fit to the observed profile by minimizing $\chi^2$.

The best fit model has  $r_{\rm core}=0\farcs 29 \pm 0\farcs 02$
and $\alpha = 0.75 \pm 0.05$,
and is shown by the red line in Fig.\ \ref{radprof.fig}.
The value of $r_{\rm max}$ is poorly constrained as there is no significant second break
in the profile at large radii; the formal fit gives $r_{\rm max} = 2\farcs 5 \pm 0\farcs 7$.
The fit is excellent, and we conclude that the surface brightness distribution
strongly indicates photoionization by a compact object that is at the
heart of a gas cloud.\footnote{In paper I we showed that there is a compact blue object coincident with
the peak of the line emission, and that its far-UV luminosity is likely sufficient to
produce the ionizing photons.}
The cloud has a constant density out to $r \approx 2.5$\,kpc from the SMBH, after which
the density decreases with radius as $\sim r^{-0.8}$.

\begin{figure*}[ht]
  \begin{center}
  \includegraphics[width=1.0\linewidth]{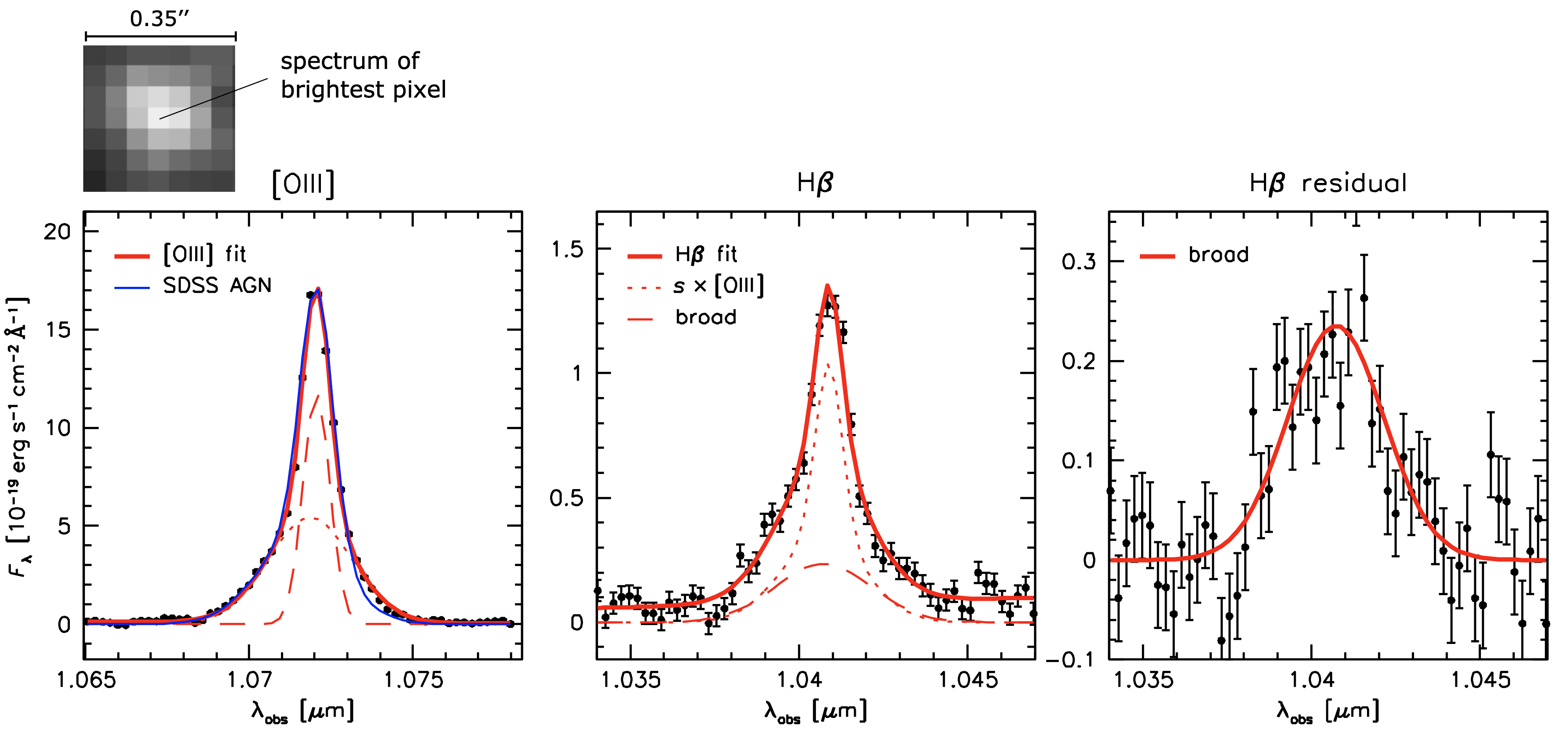}
  \end{center}
\vspace{-0.2cm}
    \caption{
Emission line fits in the central pixel. {\em Left:} The [O\,III]\,$\lambda 5007$ line.
The line profile has a blue wing, and is modeled as a combination of two Gaussians (red line).
The average [O\,III] line profile of Type 1 SDSS AGN is also shown (blue). {\em Middle:}
The H$\beta$ line is modeled with a combination of the [O\,III] profile and a broad
component, with FWHM$=970 \pm 123$\,\kms. {\em Right:} Residual of the H$\beta$ line
isolating the broad component, after
subtracting the best-fitting scaled [O\,III] profile. 
}
\label{oiii.fig}
\end{figure*}

\section{Kinematics}

Having confirmed the existence of a cloud of gas between the nuclei that is photoionized
by a SMBH, we now turn to the key question of this {\em Letter}: the radial velocity of the SMBH
with respect to the surrounding gas.

\subsection{Velocity of the SMBH}

The first task is to measure the radial velocity of the SMBH. This is not simply
the radial velocity of the gas at the location of the SMBH, as that may
largely reflect the local kinematics of the cloud, independent of the SMBH velocity.
The SMBH velocity can be inferred from gas that is either bound to it (the broad
line region, or BLR), or that is accellerated by it in an outflow.
We analyze the kinematics at the location of the SMBH in two steps, first modeling the
forbidden [O\,III] line and then
fitting for an additional broad component in H$\beta$. The H$\alpha$ line is obviously
brighter than H$\beta$, but the fit is more difficult to interpret
as H$\alpha$ is blended with the two [N\,II] lines.

The profile of the [O\,III] line in the brightest $0\farcs 05$ pixel is shown in the
left panel of Fig.\ \ref{oiii.fig}. It has a pronounced blue wing, which is very common
in AGN and usually attributed to an outflow where the far side is more obscured
than the near side \citep[e.g.,][]{mullaney:13}. For reference, the
average profile of $\sim 10,000$ Type 1 AGNs in SDSS from \citet{mullaney:13}
is shown in blue; it is a remarkably good match. We follow standard practice and fit
the line with a composite model of two Gaussians, plus a linear background.
The best-fit model is shown in red;
the two components have widths of FWHM$=272$\,\kms\ and FWHM$=840$\,\kms respectively.\footnote{A
similar asymmetric profile was seen in paper I for the [Ne\,III] line; a direct comparison is
difficult as the LRIS spectrum covers a much larger region than the single $0\farcs 05$
pixel analyzed here.}

The H$\beta$ line profile is shown in the middle panel. We fit H$\beta$ with a combination
of the [O\,III] profile, representing lower density gas, and an additional Gaussian for high
density gas where [O\,III] is suppressed.
This fit is shown by the red line. There is clearly a need for an additional broad
component, with a width of FWHM$=970 \pm 123$\,\kms. This component is isolated in the right
panel of Fig.\ \ref{oiii.fig}. This broad component could be the BLR, but it could also be turbulent
dense gas in the outflow \citep[see, e.g.,][]{harrison:14}.
The redshift of this broad component is $z_{\rm BH}=1.14023 \pm 0.00025$,
and in the following we adopt this as the radial velocity of the SMBH and the zeropoint
of the velocity field. 
We note that we
found a similar broad component for H$\gamma$ in paper I, with
FWHM$=940 \pm 110$\,\kms, with the caveat that
the red wing of the line was unconstrained due to contamination by sky lines.

\begin{figure*}[ht]
  \begin{center}
  \includegraphics[width=1.0\linewidth]{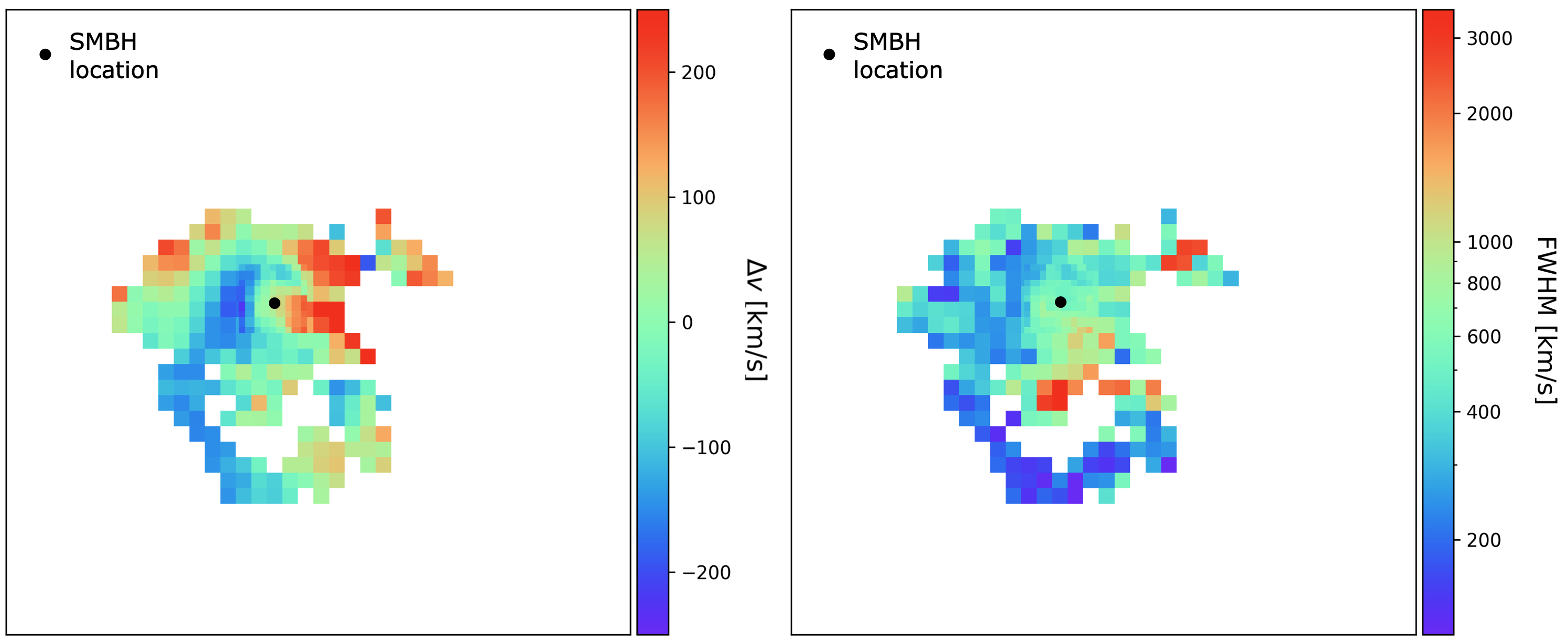}
  \end{center}
\vspace{-0.2cm}
    \caption{
{\em Left:} Velocities in the $\infty$ galaxy, with respect
to the redshift of the SMBH. Spaxels are $0\farcs 05 \times 0\farcs 05$ near the SMBH
and $0\farcs 1\times 0\farcs 1$ elsewhere.  {\em Right:} Line widths. 
The kinematics of the galaxy are complex, with strong local
velocity gradients and several regions with very broad emission lines.
The broadest lines are seen in the two nuclei and in the gas that connects them.
}
\label{vfield.fig}
\end{figure*}

\subsection{Velocity Field of the $\infty$ System}

The velocity field of the gas in the $\infty$ galaxy is determined 
in the following way.
The [O\,III] and H$\alpha$ regions are fit separately for each spaxel.
In the [O\,III] region H$\beta$ and the [O\,III]\,$\lambda\lambda 4959,5007$ doublet are fit.
Free parameters are the redshift, the width of the lines,
the flux of H$\beta$, the flux of the [O\,III] doublet (with the ratio of the two
doublet lines held fixed), and two parameters describing a linear fit to the background.
The H$\alpha$ region is fit in a similar way, except that
the velocity dispersion of H$\alpha$ is fit independently
of the dispersion of the [N\,II]\,$\lambda \lambda 6548,6583$ doublet. This is needed to fit
the lines near the two nuclei (see \S\,\ref{nuclei.sec}).
The [O\,III] and H$\alpha$ results are then combined into a single map.
The $0\farcs 05$ cube is used for the region near the SMBH, where the S/N ratio of the
cube is highest; the $0\farcs 1$ cube is used elsewhere.

The kinematics of the $\infty$ galaxy are shown in Fig.\ \ref{vfield.fig}.
The zeropoint of the velocity scale is the redshift of the SMBH, as determined above.
The velocities range from $\sim -250$\,\kms\ to $\sim +250$\,\kms, with an overall gradient
from the SE to the NW (as was also seen in the [O\,II] line in paper I). 
However, there are strong
local departures from this trend, as
expected in the aftermath of the collision that shaped the system.
Interestingly,
the strongest local gradient in the map is
from East to West across the SMBH, where the velocity changes from $-200$\,\kms\
to $+200$\,\kms\ over $\sim 6$\,kpc. This may be due to an outflow from the SMBH.
The SE ring is
mostly blueshifted by $\sim 150$\,\kms, with higher velocities in the Western part of the ring.
The two nuclei are approximately 100\,\kms\ apart.

The line widths, shown at right in Fig.\ \ref{vfield.fig}, show strong spatial variations and
complexity as well. The SMBH is in a $\sim 0\farcs 5$ diameter
region of elevated line widths with
respect to the rest of the cloud, roughly coinciding 
with the region of elevated [N\,II]/H$\alpha$ ratios in Fig.\ \ref{line_ratios.fig}
(see \S\,\ref{discussion.sec}). The smallest line widths (FWHM\,$\approx 150$\,\kms)
are seen in the SE ring, further reinforcing the interpretation that we are seeing
`regular' star formation in this area of the map.
The most striking spaxels in the line width map are
in the regions near the two nuclei
and the line connecting them, where the FWHM reaches $\sim 3000$\,\kms;
we will return to this in \S\,\ref{nuclei.sec}.

\subsection{Velocity of the SMBH With Respect to the Surrounding Gas}

\begin{figure}[ht]
  \begin{center}
  \includegraphics[width=1.0\linewidth]{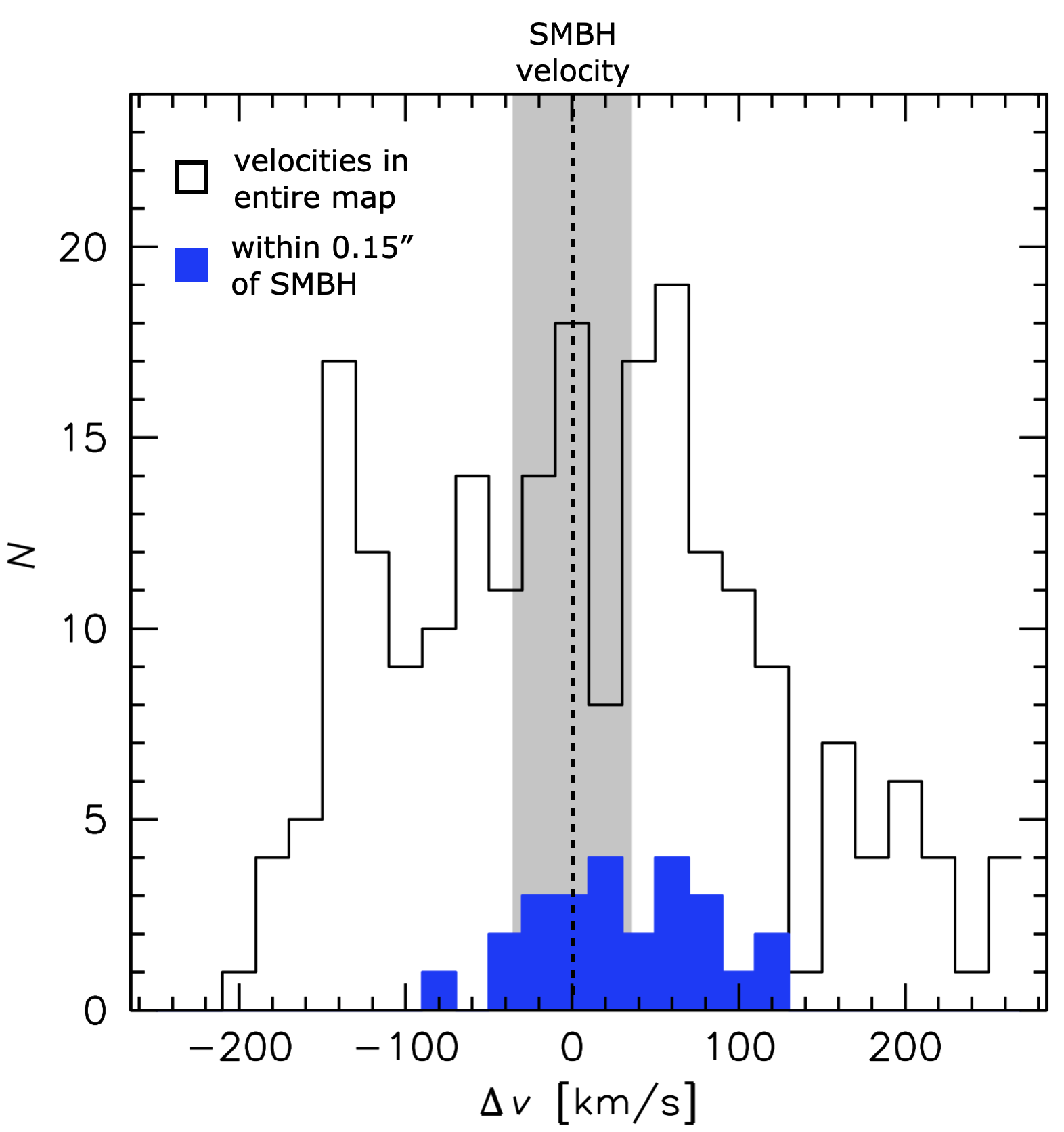}
  \end{center}
\vspace{-0.2cm}
    \caption{
Velocity distribution in the $\infty$ system. The
open histogram shows all spaxels in the $0\farcs 1$ cube;
the blue histogram shows spaxels in the $0\farcs 05$ cube that
are within a distance of $0\farcs 15$ of the SMBH. The grey band indicates the uncertainty in the redshift
of the SMBH of $\pm 35$\,\kms. The SMBH's velocity is
within $\sim 50$\,\kms\
of that of the surrounding gas.
}
\label{vhist.fig}
\end{figure}

The velocity of the SMBH is compared to the distribution of spaxel velocities
in Fig.\ \ref{vhist.fig}. The open histogram shows all velocities, measured
in $0\farcs 1 \times 0\farcs 1$ spaxels (with each spaxel having equal weight).
The offset is $3 \pm 36$\,\kms, which means that the SMBH's velocity is close
to the center of the gas velocities in the entire system. This is qualitatively
consistent with the `mini-bullet' collision scenario, where the densest
and most compressed gas is at rest with respect to the rest of the system.

\begin{figure*}[ht]
  \begin{center}
  \includegraphics[width=0.9\linewidth]{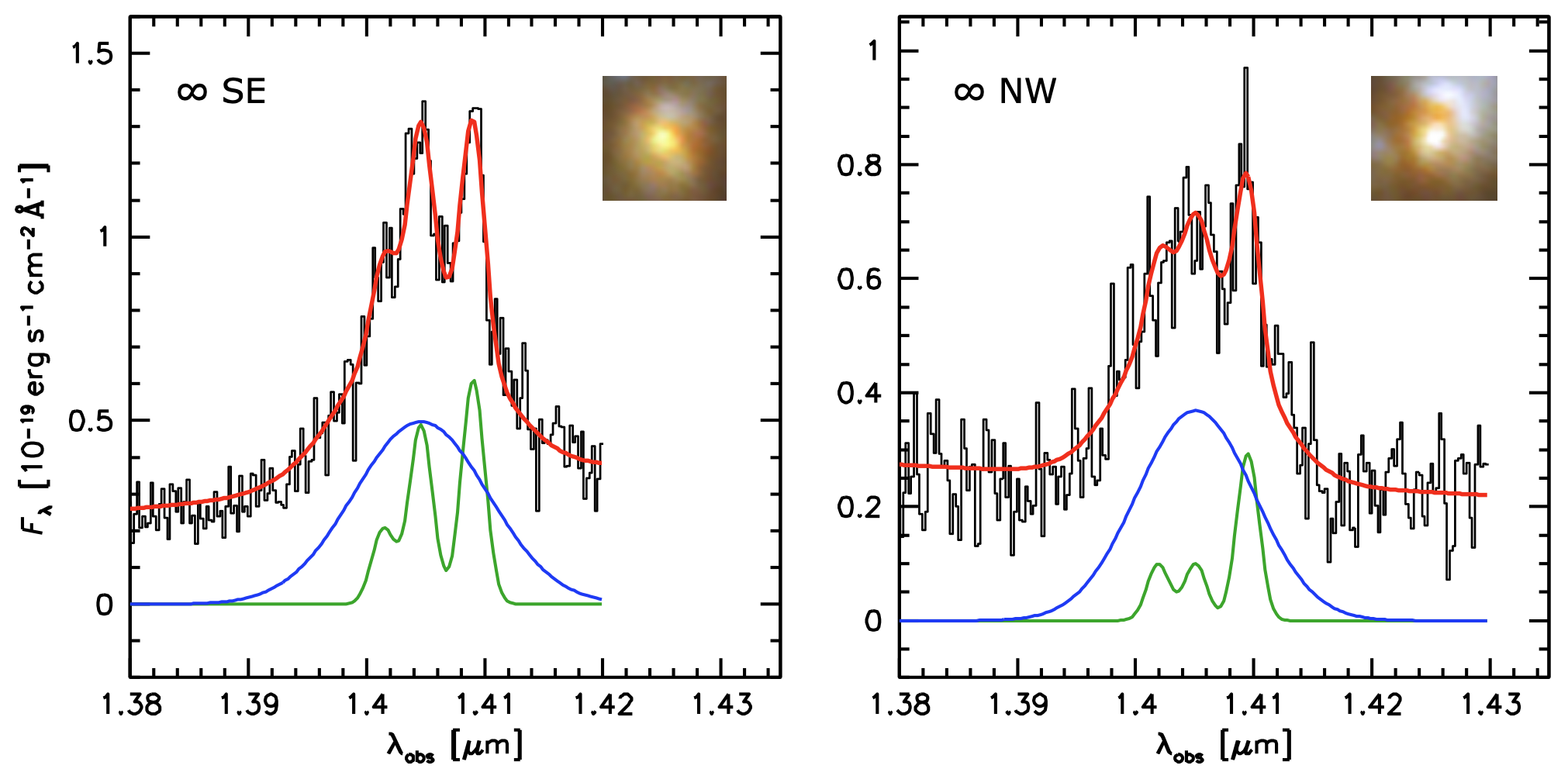}
  \end{center}
\vspace{-0.2cm}
    \caption{
Spectra of $0\farcs 2\times 0\farcs 2$ regions centered on the two nuclei.
Both nuclei show very broad H$\alpha$ emission, in addition to a narrow H$\alpha$\,+\,[N\,II] component.
The line widths of the broad components are FWHM\,$=2850$\,\kms\ and FWHM\,$=2480$\,\kms\
for $\infty$\,SE and $\infty$\,NW, respectively.
There are evidently three active SMBHs in the $\infty$ system.
}
\label{nuclei_bh.fig}
\end{figure*}

The key test is to compare the velocity of the SMBH to that of the gas in its
immediate surroundings. The blue histogram in Fig.\ \ref{vhist.fig} shows
the velocities in the 25 spaxels in the $0\farcs 05$ data cube that are
within $0\farcs 15$ (1.2\,kpc) of the SMBH.
Their mean
velocity is $31 \pm 36$\,\kms\ offset from the SMBH, and the rms scatter in these
spaxels is $\approx 50$\,\kms.
As noted in the Introduction, velocity offsets of $400$\,\kms\ -- $1500$\,\kms\ would be expected if
the SMBH came from another galaxy and is passing through the cloud.
We infer that the SMBH is, instead,
physically associated with the cloud.

\section{Two Additional Active SMBHs in the Two Nuclei}
\label{nuclei.sec}

The two massive, compact nuclei are both detected in H$\alpha$ and [N\,II], showing spatially-extended
emission with LINER-like line ratios (see Fig.\ \ref{line_ratios.fig}).
These regions have very large line widths, as shown in the right panel of Fig.\ \ref{vfield.fig}.
The spectra of the nuclear regions, summed over $0\farcs2 \times 0\farcs 2$, are shown in 
Fig.\ \ref{nuclei_bh.fig}.
We fit the spectra
with a narrow H$\alpha$\,+\,[N\,II] complex and a broad H$\alpha$ component.
For the SE nucleus, we find FWHM\,$=499 \pm 35$\,\kms\ for the narrow component and
FWHM\,$=2850\pm 200$\,\kms for the broad component. The kinematics of the NW nucleus are remarkably similar,
with FWHM$=479\pm 81$\,\kms\ and FWHM\,$=2480 \pm 170$\,\kms\ for the two components.
The velocities of the nuclei with respect to the central SMBH are $-77 \pm 37$\,\kms\ for
the SE nucleus and $+20\pm 44$\,\kms\ for the NW nucleus.

The extreme H$\alpha$ line widths, combined with the LINER line ratios, strongly indicate the presence
of a massive SMBH in each nucleus. The broad lines span kpc scales and are likely caused by
outflows \citep[see][]{harrison:14}, although emission from the BLR could contribute as well.
We also see large line widths (up to FWHM\,$\sim 1500$\,\kms) along the path connecting the two nuclei
(see Fig.\ \ref{vfield.fig}). This may be due to outflows as well, from one or both of the AGNs,
or reflect the superposition of gas that formerly belonged to the two galaxies
\citep[as seen in the Bullseye galaxy;][]{pasha:25}.

\section{Discussion}
\label{discussion.sec}

The NIRSpec IFU observations described in this {\em Letter} confirm several aspects of the 
$\infty$ galaxy. The existence of a $\sim 10$\,kpc-sized cloud of gas in between the two
nuclei is confirmed, as is the extreme equivalent width of the emission lines in
the cloud ($\sim 1600$\,\AA\ for [O\,III]\,$\lambda5007$). Furthermore, we see several
signatures of the SMBH in the heart of the cloud: the asymmetric
[O\,III] line profile that is characteristic of AGNs; the photo-ionization of
the cloud, which allows us to constrain the cloud geometry; and local effects
on the gas, seen in enhanced [N\,II]/H$\alpha$ ratios and enhanced line widths
near the SMBH. We may also see a biconal outflow in the velocity field,
to the West and East of the SMBH.

The data also strengthen the case for a collisional origin of the remarkable dual-ring morphology of the $\infty$ galaxy. The presence of AGNs in both nuclei is qualitatively
consistent with a recent collision, as simulations and observations show that such events efficiently funnel gas to the central regions \citep[see, e.g.,][]{springel:05,ellison:11}.
We may also be seeing the predicted expansion of
the SE ring \citep[see][]{lynds:76,appleton:96}
in the form of a velocity gradient, although the gradient is also consistent
with rotation.
Moreover, the turbulent gas along the path connecting the two nuclei is reminiscent of
the complex gas dynamics seen in other collisional ring galaxies \citep[e.g.,][]{appleton:96,pasha:25}.

Turning to the main goal of the IFU observations and this paper,
we find that the radial velocity of the central SMBH is only $31 \pm 36$\,\kms\
removed from that of the surrounding gas. This effectively rules out
scenarios where the SMBH escaped from one of the two nuclei.
The escape velocity
$
v_{\rm esc} \approx \left( {2GM}/a\right)^{0.5},
$
with $a=0.55 r_{\rm e}$ the scale length of a \citet{hernquist:90} profile.
Using the masses and half-light radii
of paper I, we find $v_{\rm esc} \approx 1200$\,\kms\ for the NW nucleus
and $v_{\rm esc} \approx 2700$\,\kms\ for the SE nucleus. The probability that the
line-of-sight velocity difference with the cloud is
$|\Delta v| \leq 31$\,\kms\ by chance is then 1\,\% -- 3\,\%.
We note that an escaped SMBH would have to be the result of a three-body
interaction: the
discovery of SMBHs in both nuclei rules out ejection
due to gravitational recoil following a SMBH -- SMBH merger \citep{bekenstein:73,campanelli:07},
as then one of the nuclei would no longer have a SMBH.

The observed kinematics are also difficult to reconcile with scenarios where
the SMBH is associated with a faint dwarf galaxy that is passing through the cloud. 
The 1D velocity dispersion of the $\infty$ system is approximately
$
\sigma_{\infty} \approx \left({GM}/{2R}\right)^{0.5},
$
which gives $\sigma_{\infty} \sim 350$\,\kms\ for $R \sim 5$\,kpc and
$M \sim 3\times 10^{11}$\,\msun.
Assuming that the radial velocity of the dwarf galaxy is drawn from a Gaussian distribution
with $\sigma = \sigma_{\infty}$, the probability of the observed SMBH -- cloud
velocity offset is $\sim 7$\,\%.
As discussed in paper I, the extreme equivalent widths of the emission
lines also argue against
a dwarf galaxy: they are an order of magnitude higher than the upper envelope
of AGNs in the MaNGA survey \citep{decontomachado:22}.

Instead, the small velocity difference strongly supports the idea that the central SMBH formed in situ within the cloud. This would represent the first observation of a newly formed SMBH, and constitute what is arguably the most compelling evidence yet that direct-collapse SMBH formation can occur.\footnote{We
note that the collapse
may not be a single event but could be hierarchical, with mergers of
stars leading to intermediate mass black holes that subsequently 
merged as well \citep[see, e.g.,][]{ebisuzaki:01,vergara:25}.}


The observations presented here can be extended in various ways. 
High resolution UV data would
be valuable to pinpoint the ionization source, and AO-assisted IFU
spectroscopy with 10\,m class telescope could further elucidate the gas dynamics
near the central SMBH. More generally, most of the data that
are available for the $\infty$ galaxy come from relatively shallow wide field surveys in the
COSMOS field; targeted
follow-up studies 
could provide a wealth of additional information on this unique system.

Finally, we note that the $\infty$ galaxy was independently discovered by \cite{li:25},
who dub the galaxy the Cosmic Owl. Their analysis is largely consistent with ours: the
total stellar mass for the system is very similar; they also interpret the system as a
binary collisional ring system; and they highlight the remarkable central regions of the object.
\citet{li:25} also identify the AGNs in the two nuclei,
from archival JWST NIRCAM grism data covering the Pa\,$\alpha$ line.
Besides ionized gas they find abundant molecular gas in the central cloud
from archival ALMA observations,
qualitatively consistent with the significant dust attenuation that we infer from
the Balmer decrement.

Where the studies differ is in the interpretation of the cloud in between the nuclei.
\citet{li:25} suggest that the radio source that is coincident with the peak of the
line emission (see paper I) is the lobe of a jet emanating from the NW nucleus, and that
the interaction of the lobe and the ambient gas triggered a star burst. Instead,
we associate the radio source with the central SMBH itself. Independent of the
radio data we find abundant
evidence for an active SMBH in the central gas cloud, such as the centroid of the X-ray emission,
the strong detection of [Ne\,V] and other high ionization lines from the center of
the cloud, the total [O\,III] luminosity and [O\,III]/H$\alpha$
ratios of the cloud,
the AGN-like [O\,III] line profile shown in Fig.\ \ref{oiii.fig},
and the photoionization analysis presented in \S\,\ref{photo.sec}.
We also note that the candidate second radio lobe in \citet{li:25} has low statistical
significance; deep, high spatial resolution follow-up radio observations
will be valuable in clarifying the morphology and physical nature of the
radio emission in the $\infty$ galaxy.

\begin{acknowledgements}
This paper is based on JWST data from DD program 9327.
They can be retrieved using DOI
\dataset[10.17909/gtnb-8735]{https://doi.org/10.17909/gtnb-8735}.
\end{acknowledgements}

\bibliography{master_0809}{}

\begin{thebibliography}{}
\expandafter\ifx\csname natexlab\endcsname\relax\def\natexlab#1{#1}\fi
\providecommand{\url}[1]{\href{#1}{#1}}
\providecommand{\dodoi}[1]{doi:~\href{http://doi.org/#1}{\nolinkurl{#1}}}
\providecommand{\doeprint}[1]{\href{http://ascl.net/#1}{\nolinkurl{http://ascl.net/#1}}}
\providecommand{\doarXiv}[1]{\href{https://arxiv.org/abs/#1}{\nolinkurl{https://arxiv.org/abs/#1}}}

\bibitem[{{Allen} {et~al.}(2008){Allen}, {Groves}, {Dopita}, {Sutherland}, \&
  {Kewley}}]{allen:08}
{Allen}, M.~G., {Groves}, B.~A., {Dopita}, M.~A., {Sutherland}, R.~S., \&
  {Kewley}, L.~J. 2008, \apjs, 178, 20, \dodoi{10.1086/589652}

\bibitem[{{Appleton} \& {Struck-Marcell}(1996)}]{appleton:96}
{Appleton}, P.~N., \& {Struck-Marcell}, C. 1996, \fcp, 16, 111

\bibitem[{{Appleton} {et~al.}(2022){Appleton}, {Emonts}, {Lisenfeld},
  {Falgarone}, {Guillard}, {Boulanger}, {Braine}, {Ogle}, {Struck}, {Vollmer},
  \& {Yeager}}]{appleton:22}
{Appleton}, P.~N., {Emonts}, B., {Lisenfeld}, U., {et~al.} 2022, \apj, 931,
  121, \dodoi{10.3847/1538-4357/ac63b2}

\bibitem[{{Baldwin} {et~al.}(1981){Baldwin}, {Phillips}, \&
  {Terlevich}}]{baldwin:81}
{Baldwin}, J.~A., {Phillips}, M.~M., \& {Terlevich}, R. 1981, \pasp, 93, 5

\bibitem[{{Bekenstein}(1973)}]{bekenstein:73}
{Bekenstein}, J.~D. 1973, \apj, 183, 657, \dodoi{10.1086/152255}

\bibitem[{{Brinchmann} {et~al.}(2004){Brinchmann}, {Charlot}, {White},
  {Tremonti}, {Kauffmann}, {Heckman}, \& {Brinkmann}}]{brinchmann:04}
{Brinchmann}, J., {Charlot}, S., {White}, S.~D.~M., {et~al.} 2004, \mnras, 351,
  1151, \dodoi{10.1111/j.1365-2966.2004.07881.x}

\bibitem[{{Bromm} \& {Loeb}(2003)}]{bromm:03}
{Bromm}, V., \& {Loeb}, A. 2003, \apj, 596, 34, \dodoi{10.1086/377529}

\bibitem[{Bushouse {et~al.}(2025)Bushouse, Eisenhamer, Dencheva, Davies,
  Greenfield, Morrison, Hodge, Simon, Grumm, Droettboom, Slavich, Sosey, Pauly,
  Miller, Jedrzejewski, Hack, Davis, Crawford, Law, Gordon, Regan, Cara,
  MacDonald, Bradley, Shanahan, Jamieson, Teodoro, Williams, Pena-Guerrero,
  Graham, Molter, Brandt, Hayes, Cooper, Clarke, \& Filippazzo}]{JWSTpipe}
Bushouse, H., Eisenhamer, J., Dencheva, N., {et~al.} 2025, JWST Calibration
  Pipeline,  Zenodo, \dodoi{10.5281/ZENODO.15178003}

\bibitem[{{Calzetti} {et~al.}(2000){Calzetti}, {Armus}, {Bohlin}, {Kinney},
  {Koornneef}, \& {Storchi-Bergmann}}]{calzetti:00}
{Calzetti}, D., {Armus}, L., {Bohlin}, R.~C., {et~al.} 2000, \apj, 533, 682.
\newblock
  \url{http://adsabs.harvard.edu/cgi-bin/nph-bib_query?bibcode=2000ApJ...533..682C&db_key=AST}

\bibitem[{Campanelli {et~al.}(2007)Campanelli, Lousto, Zlochower, \&
  Merritt}]{campanelli:07}
Campanelli, M., Lousto, C., Zlochower, Y., \& Merritt, D. 2007, The
  Astrophysical Journal, 659, L5, \dodoi{10.1086/516712}

\bibitem[{{Civano} {et~al.}(2010){Civano}, {Elvis}, {Lanzuisi}, {Jahnke},
  {Zamorani}, {Blecha}, {Bongiorno}, {Brusa}, {Comastri}, {Hao}, {Leauthaud},
  {Loeb}, {Mainieri}, {Piconcelli}, {Salvato}, {Scoville}, {Trump}, {Vignali},
  {Aldcroft}, {Bolzonella}, {Bressert}, {Finoguenov}, {Fruscione}, {Koekemoer},
  {Cappelluti}, {Fiore}, {Giodini}, {Gilli}, {Impey}, {Lilly}, {Lusso},
  {Puccetti}, {Silverman}, {Aussel}, {Capak}, {Frayer}, {Le Floch},
  {McCracken}, {Sanders}, {Schiminovich}, \& {Taniguchi}}]{civano:10}
{Civano}, F., {Elvis}, M., {Lanzuisi}, G., {et~al.} 2010, \apj, 717, 209,
  \dodoi{10.1088/0004-637X/717/1/209}

\bibitem[{{Clowe} {et~al.}(2006){Clowe}, {Brada{\v c}}, {Gonzalez},
  {Markevitch}, {Randall}, {Jones}, \& {Zaritsky}}]{clowe:06}
{Clowe}, D., {Brada{\v c}}, M., {Gonzalez}, A.~H., {et~al.} 2006, \apjl, 648,
  L109, \dodoi{10.1086/508162}

\bibitem[{{Deconto-Machado} {et~al.}(2022){Deconto-Machado}, {Riffel}, {Ilha},
  {Rembold}, {Storchi-Bergmann}, {Riffel}, {Schimoia}, {Schneider}, {Bizyaev},
  {Feng}, {Wylezalek}, {da Costa}, {do Nascimento}, \&
  {Maia}}]{decontomachado:22}
{Deconto-Machado}, A., {Riffel}, R.~A., {Ilha}, G.~S., {et~al.} 2022, \aap,
  659, A131, \dodoi{10.1051/0004-6361/202140613}

\bibitem[{{Dopita} \& {Sutherland}(1995)}]{dopita:95}
{Dopita}, M.~A., \& {Sutherland}, R.~S. 1995, \apj, 455, 468

\bibitem[{{Ebisuzaki} {et~al.}(2001){Ebisuzaki}, {Makino}, {Tsuru}, {Funato},
  {Portegies Zwart}, {Hut}, {McMillan}, {Matsushita}, {Matsumoto}, \&
  {Kawabe}}]{ebisuzaki:01}
{Ebisuzaki}, T., {Makino}, J., {Tsuru}, T.~G., {et~al.} 2001, \apjl, 562, L19,
  \dodoi{10.1086/338118}

\bibitem[{{Ellison} {et~al.}(2011){Ellison}, {Patton}, {Mendel}, \&
  {Scudder}}]{ellison:11}
{Ellison}, S.~L., {Patton}, D.~R., {Mendel}, J.~T., \& {Scudder}, J.~M. 2011,
  \mnras, 418, 2043, \dodoi{10.1111/j.1365-2966.2011.19624.x}

\bibitem[{{Elvis} {et~al.}(2009){Elvis}, {Civano}, {Vignali}, {Puccetti},
  {Fiore}, {Cappelluti}, {Aldcroft}, {Fruscione}, {Zamorani}, {Comastri},
  {Brusa}, {Gilli}, {Miyaji}, {Damiani}, {Koekemoer}, {Finoguenov}, {Brunner},
  {Urry}, {Silverman}, {Mainieri}, {Hasinger}, {Griffiths}, {Carollo}, {Hao},
  {Guzzo}, {Blain}, {Calzetti}, {Carilli}, {Capak}, {Ettori}, {Fabbiano},
  {Impey}, {Lilly}, {Mobasher}, {Rich}, {Salvato}, {Sanders}, {Schinnerer},
  {Scoville}, {Shopbell}, {Taylor}, {Taniguchi}, \& {Volonteri}}]{elvis:09}
{Elvis}, M., {Civano}, F., {Vignali}, C., {et~al.} 2009, \apjs, 184, 158,
  \dodoi{10.1088/0067-0049/184/1/158}

\bibitem[{{Gaskell} \& {Ferland}(1984)}]{gaskell:84}
{Gaskell}, C.~M., \& {Ferland}, G.~J. 1984, \pasp, 96, 393,
  \dodoi{10.1086/131352}

\bibitem[{{Greene} {et~al.}(2024){Greene}, {Labbe}, {Goulding}, {Furtak},
  {Chemerynska}, {Kokorev}, {Dayal}, {Volonteri}, {Williams}, {Wang}, {Setton},
  {Burgasser}, {Bezanson}, {Atek}, {Brammer}, {Cutler}, {Feldmann}, {Fujimoto},
  {Glazebrook}, {de Graaff}, {Khullar}, {Leja}, {Marchesini}, {Maseda},
  {Matthee}, {Miller}, {Naidu}, {Nanayakkara}, {Oesch}, {Pan}, {Papovich},
  {Price}, {van Dokkum}, {Weaver}, {Whitaker}, \& {Zitrin}}]{greene:24}
{Greene}, J.~E., {Labbe}, I., {Goulding}, A.~D., {et~al.} 2024, \apj, 964, 39,
  \dodoi{10.3847/1538-4357/ad1e5f}

\bibitem[{{Haehnelt} \& {Rees}(1993)}]{haehnelt:93}
{Haehnelt}, M.~G., \& {Rees}, M.~J. 1993, \mnras, 263, 168,
  \dodoi{10.1093/mnras/263.1.168}

\bibitem[{{Harrison} {et~al.}(2014){Harrison}, {Alexander}, {Mullaney}, \&
  {Swinbank}}]{harrison:14}
{Harrison}, C.~M., {Alexander}, D.~M., {Mullaney}, J.~R., \& {Swinbank}, A.~M.
  2014, \mnras, 441, 3306, \dodoi{10.1093/mnras/stu515}

\bibitem[{{Hernquist}(1990)}]{hernquist:90}
{Hernquist}, L. 1990, \apj, 356, 359, \dodoi{10.1086/168845}

\bibitem[{{Hoffman} \& {Loeb}(2007)}]{hoffman:07}
{Hoffman}, L., \& {Loeb}, A. 2007, \mnras, 377, 957,
  \dodoi{10.1111/j.1365-2966.2007.11694.x}

\bibitem[{{Kauffmann} {et~al.}(2003){Kauffmann}, {Heckman}, {Tremonti},
  {Brinchmann}, {Charlot}, \& {ETAL}}]{kauffmann:03}
{Kauffmann}, G., {Heckman}, T.~M., {Tremonti}, C., {et~al.} 2003, \mnras, 346,
  1055

\bibitem[{{Kewley} {et~al.}(2013){Kewley}, {Dopita}, {Leitherer}, {Dav{\'e}},
  {Yuan}, {Allen}, {Groves}, \& {Sutherland}}]{kewley:13}
{Kewley}, L.~J., {Dopita}, M.~A., {Leitherer}, C., {et~al.} 2013, \apj, 774,
  100, \dodoi{10.1088/0004-637X/774/2/100}

\bibitem[{{Lee} {et~al.}(2021){Lee}, {Shin}, \& {Kim}}]{lee:21}
{Lee}, J., {Shin}, E.-j., \& {Kim}, J.-h. 2021, \apjl, 917, L15,
  \dodoi{10.3847/2041-8213/ac16e0}

\bibitem[{{Li} {et~al.}(2025){Li}, {Emonts}, {Cai}, {Tanaka}, {Mercier}, {Wu},
  {Yu}, {Sun}, {Bian}, {Daddi}, {Fan}, {Lin}, {Lyu}, {Kartaltepe}, \&
  {Valentino}}]{li:25}
{Li}, M., {Emonts}, B. H.~C., {Cai}, Z., {et~al.} 2025, arXiv e-prints,
  arXiv:2506.10058, \dodoi{10.48550/arXiv.2506.10058}

\bibitem[{{Lodato} \& {Natarajan}(2006)}]{lodato:06}
{Lodato}, G., \& {Natarajan}, P. 2006, \mnras, 371, 1813,
  \dodoi{10.1111/j.1365-2966.2006.10801.x}

\bibitem[{{Lousto} \& {Zlochower}(2011)}]{lousto:11}
{Lousto}, C.~O., \& {Zlochower}, Y. 2011, \prd, 83, 024003,
  \dodoi{10.1103/PhysRevD.83.024003}

\bibitem[{{Lynds} \& {Toomre}(1976)}]{lynds:76}
{Lynds}, R., \& {Toomre}, A. 1976, \apj, 209, 382, \dodoi{10.1086/154730}

\bibitem[{{Madau} \& {Rees}(2001)}]{madau:01}
{Madau}, P., \& {Rees}, M.~J. 2001, \apjl, 551, L27, \dodoi{10.1086/319848}

\bibitem[{{Magorrian} {et~al.}(1998){Magorrian}, {Tremaine}, {Richstone},
  {Bender}, {Bower}, {Dressler}, {Faber}, {Gebhardt}, {Green}, {Grillmair},
  {Kormendy}, \& {Lauer}}]{magorrian:98}
{Magorrian}, J., {Tremaine}, S., {Richstone}, D., {et~al.} 1998, \aj, 115, 2285

\bibitem[{{Matthee} {et~al.}(2024){Matthee}, {Naidu}, {Brammer}, {Chisholm},
  {Eilers}, {Goulding}, {Greene}, {Kashino}, {Labbe}, {Lilly}, {Mackenzie},
  {Oesch}, {Weibel}, {Wuyts}, {Xiao}, {Bordoloi}, {Bouwens}, {van Dokkum},
  {Illingworth}, {Kramarenko}, {Maseda}, {Mason}, {Meyer}, {Nelson}, {Reddy},
  {Shivaei}, {Simcoe}, \& {Yue}}]{matthee:24}
{Matthee}, J., {Naidu}, R.~P., {Brammer}, G., {et~al.} 2024, \apj, 963, 129,
  \dodoi{10.3847/1538-4357/ad2345}

\bibitem[{{Mayer} {et~al.}(2015){Mayer}, {Fiacconi}, {Bonoli}, {Quinn},
  {Ro{\v{s}}kar}, {Shen}, \& {Wadsley}}]{mayer:15}
{Mayer}, L., {Fiacconi}, D., {Bonoli}, S., {et~al.} 2015, \apj, 810, 51,
  \dodoi{10.1088/0004-637X/810/1/51}

\bibitem[{{Mayer} {et~al.}(2010){Mayer}, {Kazantzidis}, {Escala}, \&
  {Callegari}}]{mayer:10}
{Mayer}, L., {Kazantzidis}, S., {Escala}, A., \& {Callegari}, S. 2010, \nat,
  466, 1082, \dodoi{10.1038/nature09294}

\bibitem[{{Mullaney} {et~al.}(2013){Mullaney}, {Alexander}, {Fine}, {Goulding},
  {Harrison}, \& {Hickox}}]{mullaney:13}
{Mullaney}, J.~R., {Alexander}, D.~M., {Fine}, S., {et~al.} 2013, \mnras, 433,
  622, \dodoi{10.1093/mnras/stt751}

\bibitem[{{Natarajan} {et~al.}(2017){Natarajan}, {Pacucci}, {Ferrara},
  {Agarwal}, {Ricarte}, {Zackrisson}, \& {Cappelluti}}]{natarajan:17}
{Natarajan}, P., {Pacucci}, F., {Ferrara}, A., {et~al.} 2017, \apj, 838, 117,
  \dodoi{10.3847/1538-4357/aa6330}

\bibitem[{{Natarajan} {et~al.}(2024){Natarajan}, {Pacucci}, {Ricarte},
  {Bogd{\'a}n}, {Goulding}, \& {Cappelluti}}]{natarajan:24}
{Natarajan}, P., {Pacucci}, F., {Ricarte}, A., {et~al.} 2024, \apjl, 960, L1,
  \dodoi{10.3847/2041-8213/ad0e76}

\bibitem[{{Pasha} {et~al.}(2025){Pasha}, {van Dokkum}, {Liu}, {Bowman},
  {Janssens}, {Keim}, {Neufeld}, \& {Abraham}}]{pasha:25}
{Pasha}, I., {van Dokkum}, P.~G., {Liu}, Q., {et~al.} 2025, \apjl, 980, L3,
  \dodoi{10.3847/2041-8213/ad9f5c}

\bibitem[{{Privon} {et~al.}(2008){Privon}, {O'Dea}, {Baum}, {Axon}, {Kharb},
  {Buchanan}, {Sparks}, \& {Chiaberge}}]{privon:08}
{Privon}, G.~C., {O'Dea}, C.~P., {Baum}, S.~A., {et~al.} 2008, \apjs, 175, 423,
  \dodoi{10.1086/525024}

\bibitem[{{Rauscher}(2024)}]{rauscher:24}
{Rauscher}, B.~J. 2024, \pasp, 136, 015001, \dodoi{10.1088/1538-3873/ad1b36}

\bibitem[{{Regan} \& {Volonteri}(2024)}]{regan:24}
{Regan}, J., \& {Volonteri}, M. 2024, The Open Journal of Astrophysics, 7, 72,
  \dodoi{10.33232/001c.123239}

\bibitem[{Rigby {et~al.}(2023)Rigby, Vieira, Phadke, Hutchison, Welch, Cathey,
  Spilker, Gonzalez, Adhikari, Aravena, Bayliss, Birkin, Bursk, Chapman, Dahle,
  Elicker, Fischer, Florian, Gladders, Hayward, Hewald, Kettler, Khullar, Kim,
  Law, Mahler, Malhotra, Murphy, Narayanan, Olivier, Rhoads, Sharon, Solimano,
  Thiruvengadam, Vizgan, \& Younker}]{TEMPLATES}
Rigby, J.~R., Vieira, J.~D., Phadke, K.~A., {et~al.} 2023, JWST Early Release
  Science Program TEMPLATES: Targeting Extremely Magnified Panchromatic Lensed
  Arcs and their Extended Star formation.
\newblock \doarXiv{2312.10465}

\bibitem[{{Saslaw} {et~al.}(1974){Saslaw}, {Valtonen}, \&
  {Aarseth}}]{saslaw:74}
{Saslaw}, W.~C., {Valtonen}, M.~J., \& {Aarseth}, S.~J. 1974, \apj, 190, 253,
  \dodoi{10.1086/152870}

\bibitem[{{Silk}(2019)}]{silk:19}
{Silk}, J. 2019, \mnras, 488, L24, \dodoi{10.1093/mnrasl/slz090}

\bibitem[{{Springel} {et~al.}(2005){Springel}, {Di Matteo}, \&
  {Hernquist}}]{springel:05}
{Springel}, V., {Di Matteo}, T., \& {Hernquist}, L. 2005, \mnras, 361, 776,
  \dodoi{10.1111/j.1365-2966.2005.09238.x}

\bibitem[{{Tilak} {et~al.}(2005){Tilak}, {O'Dea}, {Tadhunter}, {Wills},
  {Morganti}, {Baum}, {Koekemoer}, \& {Dallacasa}}]{tilak:05}
{Tilak}, A., {O'Dea}, C.~P., {Tadhunter}, C., {et~al.} 2005, \aj, 130, 2513,
  \dodoi{10.1086/497265}

\bibitem[{{Tremmel} {et~al.}(2018){Tremmel}, {Governato}, {Volonteri},
  {Pontzen}, \& {Quinn}}]{tremmel:18}
{Tremmel}, M., {Governato}, F., {Volonteri}, M., {Pontzen}, A., \& {Quinn},
  T.~R. 2018, \apjl, 857, L22, \dodoi{10.3847/2041-8213/aabc0a}

\bibitem[{{Uppal} {et~al.}(2024){Uppal}, {Ward}, {Gezari}, {Natarajan}, {Chen},
  {LaChance}, \& {Di Matteo}}]{uppal:24}
{Uppal}, A., {Ward}, C., {Gezari}, S., {et~al.} 2024, \apj, 975, 286,
  \dodoi{10.3847/1538-4357/ad7ff0}

\bibitem[{{van Dokkum} {et~al.}(2025){van Dokkum}, {Brammer}, {Baggen}, {Keim},
  {Natarajan}, \& {Pasha}}]{dokkum:25a}
{van Dokkum}, P., {Brammer}, G., {Baggen}, J. F.~W., {et~al.} 2025, arXiv
  e-prints, arXiv:2506.15618, \dodoi{10.48550/arXiv.2506.15618}

\bibitem[{{van Dokkum} {et~al.}(2023){van Dokkum}, {Pasha}, {Buzzo}, {LaMassa},
  {Shen}, {Keim}, {Abraham}, {Conroy}, {Danieli}, {Mitra}, {Nagai},
  {Natarajan}, {Romanowsky}, {Tremblay}, {Urry}, \& {van den
  Bosch}}]{dokkum:23}
{van Dokkum}, P., {Pasha}, I., {Buzzo}, M.~L., {et~al.} 2023, \apjl, 946, L50,
  \dodoi{10.3847/2041-8213/acba86}

\bibitem[{{Vergara} {et~al.}(2025){Vergara}, {Askar}, {Kamlah}, {Spurzem},
  {Flammini Dotti}, {Schleicher}, {Arca Sedda}, {Hypki}, {Giersz}, {Hurley},
  {Berczik}, {Escala}, {Hoyer}, {Neumayer}, {Pang}, {Tanikawa}, {Cen}, \&
  {Naab}}]{vergara:25}
{Vergara}, M.~C., {Askar}, A., {Kamlah}, A. W.~H., {et~al.} 2025, arXiv
  e-prints, arXiv:2505.07491, \dodoi{10.48550/arXiv.2505.07491}

\bibitem[{{Volonteri}(2010)}]{volonteri:10}
{Volonteri}, M. 2010, \aapr, 18, 279, \dodoi{10.1007/s00159-010-0029-x}

\bibitem[{{Wise} {et~al.}(2019){Wise}, {Regan}, {O'Shea}, {Norman}, {Downes},
  \& {Xu}}]{wise:19}
{Wise}, J.~H., {Regan}, J.~A., {O'Shea}, B.~W., {et~al.} 2019, \nat, 566, 85,
  \dodoi{10.1038/s41586-019-0873-4}

\bibitem[{{Yeager} \& {Struck}(2019)}]{yeager:19}
{Yeager}, T.~R., \& {Struck}, C. 2019, \mnras, 486, 2660,
  \dodoi{10.1093/mnras/stz916}

\end{thebibliography}
\bibliographystyle{aasjournal}

\end{document}